\documentclass[twocolumn,showpacs,amsmath,amssymb]{revtex4}

\usepackage{graphicx}
\usepackage{dcolumn}
\usepackage{bm}

\begin{document}

\title{Deformation effects on the Gamow-Teller strength distributions in the
double-beta decay partners $^{76}$Ge and $^{76}$Se}

\author{P.~Sarriguren}
\affiliation{Instituto de Estructura de la Materia, IEM-CSIC, Serrano
123, E-28006 Madrid, Spain} 
\date{\today}

\begin{abstract}

A theoretical approach based on a deformed quasiparticle random phase
approximation built on a Skyrme selfconsistent mean field is used to
describe the recent measurements of the Gamow-Teller GT$^-$ strength
distribution extracted from the charge-exchange reaction
$^{76}$Ge($^3$He$,t)^{76}$As with high energy resolution. 
The same analysis is made to describe the Gamow-Teller GT$^+$ strength
distribution measured in the $^{76}$Se$(d,^2$He)$^{76}$As reaction. 
Combining these two branches, the nuclear matrix element for the 
two-neutrino double-beta decay process is evaluated and compared to 
experiment. The role of the nuclear deformation on those processes is
emphasized and analyzed.

\end{abstract}


\maketitle

\section{Introduction\label{sec:introduction}}

The Gamow-Teller (GT) nuclear response is a very fertile source of
information about important issues related not only to nuclear physics 
\cite{fujita11}, but also to astrophysics \cite{langanke03} and 
particle physics \cite{2breview,hardy}.
In the case of unstable nuclei this information is mainly extracted from
$\beta$ decays, where there is a severe restriction due to the $Q$-energy
limitation. In the case of stable or close to stability nuclei, the GT
strength is obtained from charge-exchange reactions at intermediate incident
energies and forward angles \cite{osterfeld92}. Under these conditions
the nuclear states are probed at small momentum transfer and the cross
section becomes proportional to the GT matrix element without the energy
limitations that characterize $\beta$ decays.

The spin-isospin nuclear properties in $^{76}$Ge and $^{76}$Se are among
the most extensively studied both theoretically
\cite{muto88,sahu99,sarri03,simkovic04,suh08,yousef09,moreno10} and 
experimentally \cite{madey89,thies12,helmer97,grewe08,sch08,kay09}. 
This is due to their significance as double-$\beta$ decay partners and
the implications of this process in the determination of the neutrino
nature and its absolute mass \cite{2breview}.
The study of these nuclei is indeed a part of a large experimental 
program being pursued in the last several years and aimed to explore the GT
properties at low excitation energies of double-$\beta$ decay partners
by high resolution charge-exchange 
reactions \cite{fujita11,grewe08,frekers,frekers2,zegers07}. 

The present work is motivated by the recent high resolution charge-exchange
experiment $^{76}$Ge($^3$He$,t)^{76}$As \cite{thies12} that has allowed the
unveilling of some remarkable features of this nucleus, which previous
charge-exchange experiments \cite{madey89} at much lower resolution
were unable to identify. In particular the authors of Ref.~\cite{thies12} 
reported an unusually strong fragmentation of the GT strength that was 
interpreted in terms of possible effects of deformation. In 
Ref.~\cite{thies12} it was also noted a lack of correlation among the
GT transition strengths feeding the same levels in $^{76}$As from the
two different directions, GT$^-$ measured in the $^{76}$Ge($^3$He$,t)^{76}$As 
reaction and GT$^+$ measured in the $^{76}$Se$(d,^2$He)$^{76}$As 
reaction \cite{grewe08}. In view of this new experimental information that
has become available, it is worth reconsidering the theoretical description
of these nuclei and the role that deformation might play to understand
the observed features.

In this work we explore the ability of the deformed proton-neutron 
quasiparticle random phase approximation (QRPA) approach to describe
together all of this rich information available at present that includes 
i) the global properties of the GT response, such as the total GT strength
as well as the location and strength of the GT resonance, ii) the GT
strength distribution in the low-lying excitation region that contains
much more accurate information, and iii) the two-neutrino double-$\beta$ 
($2\nu\beta\beta$) decay matrix element and the implications of the
single $\beta$ branches in the $2\nu\beta\beta$ process.

The QRPA is one of the most reliable and broadly used microscopic
approximations for calculating the correlated wave functions involved
in $\beta$ and double $\beta$ \cite{2bqrpa} decay processes, especially
after the inclusion of particle-hole ($ph$) and particle-particle ($pp$)
residual interactions. The method was first studied in Ref.~\cite{halb}
to describe the $\beta$ strength in spherical nuclei. Subsequent
extensions of the QRPA method to deal with deformed nuclei were done
later in Refs.~\cite{kru,moller,hir,hamamoto,sarriguren,sarri_pp}. Deformation
effects were also studied in the double-$\beta$ decay process
\cite{sahu99,simkovic04,alvarez04,chandra05,singh07,yousef09}.
In particular, it has been found  \cite{simkovic04,alvarez04} that the
nuclear matrix elements for the $2\nu\beta\beta$ process are suppressed
with respect to the spherical case with a reduction factor that scales
roughly with the deformation difference between parent and daughter. 
This suppression mechanism, which is ignored in spherical treatments
may play an important role in approaching the theoretical estimates
to the experiment.

As it shall be described later, our theoretical approach \cite{sarriguren,sarri_pp}
is based on a deformed QRPA formalism on top of a selfconsistent deformed 
Hartree-Fock (HF) mean field with Skyrme forces and pairing correlations 
in the BCS approximation. 
In particular, we shall study the dependence on deformation of the single
$\beta$ branches that build up the double-$\beta$ process.

The paper is organized as follows.
In Sec. II, we present a brief summary of the theoretical approach used
to describe the GT properties. Section III contains the results obtained 
for the GT strength distributions, as well as the results for the
$2\nu\beta\beta$ decay, stressing the effect of deformation. 
The summary and conclusions are given in Sec. IV.

\section{Theoretical approach}
\label{sec:theory}

We describe here briefly the theoretical formalism used in this work, whose details
can be found in Ref.~\cite{sarriguren,sarri_pp}. 
First, we carry out a selfconsistent deformed
HF calculation with the effective nucleon-nucleon density-dependent
Skyrme interaction SLy4 \cite{chabanat98}, assuming axial deformation and
time reversal symmetry \cite{vautherin}. The single-particle wave
functions are expanded in terms of the eigenstates of an axially symmetric
harmonic oscillator in cylindrical coordinates using twelve major shells.
Pairing correlations between like nucleons are included in the BCS approximation
taking fixed pairing gap parameters for protons and neutrons, which are
determined phenomenologically from the odd-even mass differences of
neighboring nuclei through a symmetric five-term formula involving
experimental binding energies \cite{audi03}. 
The occupation probabilities $v_i^2$ of the single particle levels are 
computed at the end of each HF iteration and are then used
to calculate the one-body density and mean field of the next iteration, 
so that one gets new single-particle wave functions, energies and occupation 
numbers at each iteration. Therefore,
the selfconsistent determination of the binding energy and deformation
includes pairing correlations from the beginning. After convergence, 
the QRPA equations are solved on the deformed ground state basis for 
$^{76}$Ge and $^{76}$Se to get their GT strength distributions and to 
compute the $2\nu\beta\beta$ decay matrix element.

To describe GT excitations in the QRPA we add to the quasiparticle mean
field a separable spin-isospin residual interaction in $ph$ and 
$pp$ channels. The advantage of using
separable forces is that the QRPA energy eigenvalue problem is reduced to
finding the roots of an algebraic equation. The $ph$ part is responsible for 
the position and structure of the GT resonance \cite{sarriguren,moller,hir}. 
Its coupling constant $\chi_{ph}^{GT}$ could in principle be obtained in a 
consistent way from the same Skyrme energy density functional as the HF
mean field through the second derivatives of the energy functional with
respect to the densities and averaging the contact interaction over the
nuclear volume, as it was done in Ref.~\cite{sarriguren} to study exotic
nuclei. The $pp$ part consists of a proton-neutron pairing force and it
is also introduced as a separable force \cite{hir}. The coupling constant 
$\kappa_{pp}^{GT}$ is usually fitted to the half-lives phenomenology 
\cite{hir}. Following the above mentioned procedure and taking into
account the experience accumulated in this mass region \cite{sarri09},
we have chosen in this work the values  $\chi_{ph}^{GT}=0.15$ MeV and
$\kappa_{pp}^{GT}=0.03$ MeV. In addition, we will also show the sensitivity
of the GT strength distributions and $2\nu\beta\beta$ nuclear matrix
elements to the value of coupling constant $\kappa_{pp}^{GT}$.

The technical details to solve the QRPA equations have been described in
Refs.~\cite{hir,sarriguren,sarri_pp}. Here we only mention that, because of the
use of separable residual forces, the solutions of the QRPA equations are
found by solving first a dispersion relation, which is an algebraic
equation of fourth order in the excitation energy $\omega$. Then, for
each value of the energy, the GT transition amplitudes in the intrinsic
frame connecting the ground state $\left| 0\right\rangle $ to one phonon
states in the daughter nucleus $\left| \omega _K \right\rangle $, are 
found to be

\begin{equation} 
\left\langle \omega _K | \sigma _Kt^{\pm} | 0 \right\rangle = \mp 
M^{\omega _K}_\pm \, ,
\label{amplgt}
\end{equation}
where 
\begin{eqnarray}
M_{-}^{\omega _{K}}&=&\sum_{\pi\nu}\left( q_{\pi\nu}X_{\pi\nu}^{\omega _{K}}+
\tilde{q}_{\pi\nu}Y_{\pi\nu}^{\omega _{K}}\right) , \\ 
M_{+}^{\omega _{K}}&=&\sum_{\pi\nu}\left( \tilde{q}_{\pi\nu}
X_{\pi\nu}^{\omega _{K}}+ q_{\pi\nu}Y_{\pi\nu}^{\omega _{K}}\right) \, ,
\end{eqnarray}
with 
\begin{equation}
\tilde{q}_{\pi\nu}=u_{\nu}v_{\pi}\Sigma _{K}^{\nu\pi }\, ,\ \ 
q_{\pi\nu}=v_{\nu}u_{\pi}\Sigma _{K}^{\nu\pi}\, ,\ \ 
\Sigma _{K}^{\nu\pi}=\left\langle \nu\left| \sigma _{K}\right| 
\pi\right\rangle \, ,
\label{qs}
\end{equation}
in terms of the occupation amplitudes for neutrons and protons
$v_{\nu,\pi}$ ($u^2_{\nu,\pi}=1-v^2_{\nu,\pi}$)
and the matrix elements of the spin operator connecting
proton and neutron single-particle states, as they come out from the 
HF+BCS calculation.
$X_{\pi\nu}^{\omega _{K}}$ and $Y_{\pi\nu}^{\omega _{K}}$ are the forward and 
backward amplitudes of the QRPA phonon operator, respectively.

Once the intrinsic amplitudes in Eq.~(\ref{amplgt}) are calculated, the
Gamow-Teller strength $B$(GT) in the laboratory frame for a transition 
$I_i K_i (0^+0)\rightarrow I_fK_f(1^+K)$ can be obtained as

\begin{eqnarray}
B_{\omega}({\rm GT}^\pm ) &=& \sum_{\omega_{K}} \left[ \left\langle \omega_{K=0} 
\left| \sigma_0t^\pm \right| 0 \right\rangle ^2 \delta (\omega_{K=0}-
\omega ) \right. \nonumber \\
&& + 2\left. \left\langle \omega_{K=1} \left| \sigma_1t^\pm \right| 
0 \right\rangle ^2 \delta (\omega_{K=1}-\omega ) \right] \, ,
\label{bgt}
\end{eqnarray}
in [$g_A^2/4\pi$] units. 
To obtain this expression we have used the initial and final states in 
the laboratory frame expressed in terms of the intrinsic states using 
the Bohr and Mottelson factorization \cite{bm}. Finally, a quenching factor 
$g_{A,{\rm eff}}=0.7\ g_{A,{\rm free}}$ is included in the calculations to 
take into account in an effective way all the correlations \cite{hama_eff} 
that are not properly considered in the present approach.

The role of the residual interactions and BCS correlations on the GT
strengths was already studied in Ref.~\cite{sarriguren,sarri_pp}. The role of
deformation was also studied there, where it was
shown that the GT strength distributions corresponding to deformed nuclei
are much more fragmented than the corresponding to spherical ones, because
of the broken degeneracy of the spherical shells. It was also shown that
the crossing of deformed energy levels, which depends on the magnitude of
the quadrupole deformation as well as on the oblate or prolate character, 
may lead to sizable differences between the GT strength distributions 
corresponding to different shapes.
These features have been exploited to use the $\beta$-decay properties 
as an alternative method to learn about the nuclear deformation in highly
unstable isotopes \cite{nacher}.
It has also been shown \cite{moreno09} that deformation is a key ingredient
to reproduce the occupation probabilities of the relevant single particle
levels in the valence shells of $^{76}$Ge and $^{76}$Se involved in the 
double-beta decay process that have been measured for neutrons \cite{sch08} 
and protons \cite{kay09}. 

The nuclear double-$\beta$ decay is a rare second order weak interaction
process that takes place when the transition to the intermediate nucleus
is energetically forbidden or highly retarded. Two decay modes are expected,
the two neutrino mode, involving the emission of two electrons and 
two neutrinos, and the neutrinoless mode with no neutrino leaving the 
nucleus. Whereas the first type is perfectly compatible with the Standard 
Model, the second one violates lepton number conservation and implies
the existence of a massive Majorana neutrino. Because the nuclear wave
functions and the underlying theory for treating the neutrinoless and the
two-neutrino modes are similar, the Gamow-Teller part that drives the 
$2\nu\beta\beta$ decay provides insight for theoretical models that are
required to reproduce the available experimental information on the 
$2\nu\beta\beta$ half-lives.

The $2\nu\beta\beta$ decay is described in the second order perturbation of the 
weak interaction as two successive Gamow-Teller transitions via virtual
intermediate $1^+$ states. The basic expressions for the $2\nu\beta\beta$ 
decay within a deformed QRPA formalism can be found in 
\cite{simkovic04,moreno09}. Here we only write the half-life of the 
$2\nu\beta\beta$ decay 

\begin{equation}\label{half-life}
\left[ T_{1/2}^{2\nu\beta\beta}\left( 0^+_{\rm gs} \to 0^+_{\rm gs}  
\right) \right] ^{-1}= (g_A)^4\ G^{2\nu\beta\beta}\ \left| 
M_{GT}^{2\nu\beta\beta}\right| ^2 \, ,
\end{equation}
in terms of the phase-space integral $G^{2\nu\beta\beta}$ and the nuclear 
matrix element $M_{GT}^{2\nu\beta\beta}$ that contains all the 
information of the nuclear structure involved in the process,

\begin{eqnarray}\label{mgt}
&& M_{GT}^{2\nu\beta\beta}=\sum_{K=0,\pm 1}\sum_{m_i,m_f} (-1)^K 
\frac{\langle \omega_{K,m_f}  | 
\omega_{K,m_i}  \rangle }
{(\omega_K^{m_f}+ \omega_K^{m_i}) / 2} \nonumber \\
&& \times \langle 0_f| \sigma_{-K}t^-| \omega_{K,m_f}  \rangle \: 
\langle \omega_{K,m_i}  | \sigma_Kt^- | 0_i \rangle \, .
\end{eqnarray}
In this equation $\omega_K^{m_i} (\omega_K^{m_f})$ are the QRPA excitation
energies of the intermediate $1^+$ states 
$|\omega_{K,m_i}\rangle (|\omega_{K,m_f}\rangle)$ with respect to the
initial (final) nucleus. The indices $m_i$, $m_f$ label the $1^+$ states
of the intermediate nucleus. The overlaps are needed to take into account
the non-orthogonality of the intermediate states reached from different 
initial $| 0_i \rangle$ and final $ | 0_f \rangle$ ground states. 
Their expressions can be found in Ref.~\cite{simkovic04}.

The various measurements reported for the  $2\nu\beta\beta$ decay in 
$^{76}$Ge have been analyzed in Ref.~\cite{barabash10}, where a
recommended value $ T_{1/2}^{2\nu\beta\beta} =(1.5 \pm 0.1)\times 10^{21}$ yr was 
adopted. Using the phase-space factor  \cite{simkovic04}
$G^{2\nu\beta\beta}\ (^{76}$Ge)$ = 1.49 10^{-20}$ yr$^{-1}$ MeV$^2$, we get the 
experimental nuclear matrix elements $M_{GT}^{2\nu\beta\beta}=0.129$ MeV$^{-1}$ 
when the bare $g_A=1.269$ is used and $M_{GT}^{2\nu\beta\beta}=0.216$ MeV$^{-1}$ 
when quenched factors $g_A=1$ are used.

\section{Results and Discussion\label{sec:discussion}}

\begin{table}[ctb]
\begin{center}
\caption{
Experimental and calculated charge root mean square radii $r_c$ [fm], 
intrinsic charge quadrupole moments $Q_p$ [fm$^2$], and quadrupole 
deformations $\beta$ for $^{76}$Ge and $^{76}$Se. Experimental values for 
$r_c$ are from \cite{radiiexp}. The first experimental values for $Q_p$ 
are from \cite{stone05}, while the second values are from \cite{raman01}.}
\label{rqb}
\vskip 0.5cm

\begin{tabular}{llccc}
& & $r_c$ & $Q_p$ & $\beta$ \\ 
\hline \\
$^{76}$Ge & exp. & 4.080/4.127 & 66(21)/164.1(2.5) & 0.10/0.26 \\ 
          & SLy4  & 4.104 & 93.85 & 0.14   \\ \\
$^{76}$Se & exp. & 4.088/4.162 & 119(25)/205.5(2.4) & 0.16/0.31 \\ 
          & SLy4  & 4.151 & 125.1 & 0.17  \\
\hline
\end{tabular}
\end{center}
\end{table}

\begin{table*}[ctb]
\begin{center}
\caption{Measured and calculated GT strength in $^{76}$Ge 
accumulated in various energy regions [MeV].}
\label{gtge}
\vskip 0.5cm

\begin{tabular}{lccccc}
$\sum B({\rm GT}^-)$ & $\sum_{0-5}$ & $\sum_{0-7}$ & $\sum_{7-10}$  &  
$\sum_{10-20}$ &  $\sum_{0-20}$ \\ 
\hline \\
Thies et al. \cite{thies12}  & 1.60(18) / 2.43(32) & & & & \\ 
Madey et al. \cite{madey89}  & 1.52 & 4.88 & 2.58 & 12.43 & 19.9 \\ 
QRPA ($\beta=0.15$) & 4.0 & 5.5 & 2.2 & 10.6 & 18.3 \\
\hline
\end{tabular}
\end{center}
\end{table*}

\begin{table*}[ctb]
\begin{center}
\caption{Same as in Table~\ref{gtge}, but for $^{76}$Se.} 
\label{gtse}
\vskip 0.5cm

\begin{tabular}{lcccc}
$\sum B({\rm GT}^+)$ & $\sum_{0-2}$ & $\sum_{0-5}$ & $\sum_{5-20}$ & 
$\sum_{0-20}$ \\ 
\hline \\
Grewe et al. \cite{grewe08}  & 0.54(7) & 0.7(2) & 0.3(2) & 1.0(4) \\ 
Helmer et al. \cite{helmer97}  & 0.38(3) & 0.79(7) & 0.64(9) & 1.43(16) \\ 
QRPA ($\beta=0.2$) & 0.15 & 0.87 & 0.53 & 1.4 \\
\hline
\end{tabular}
\end{center}
\end{table*}

\begin{figure}[ctb]
\begin{center}
\includegraphics[width=8.5cm]{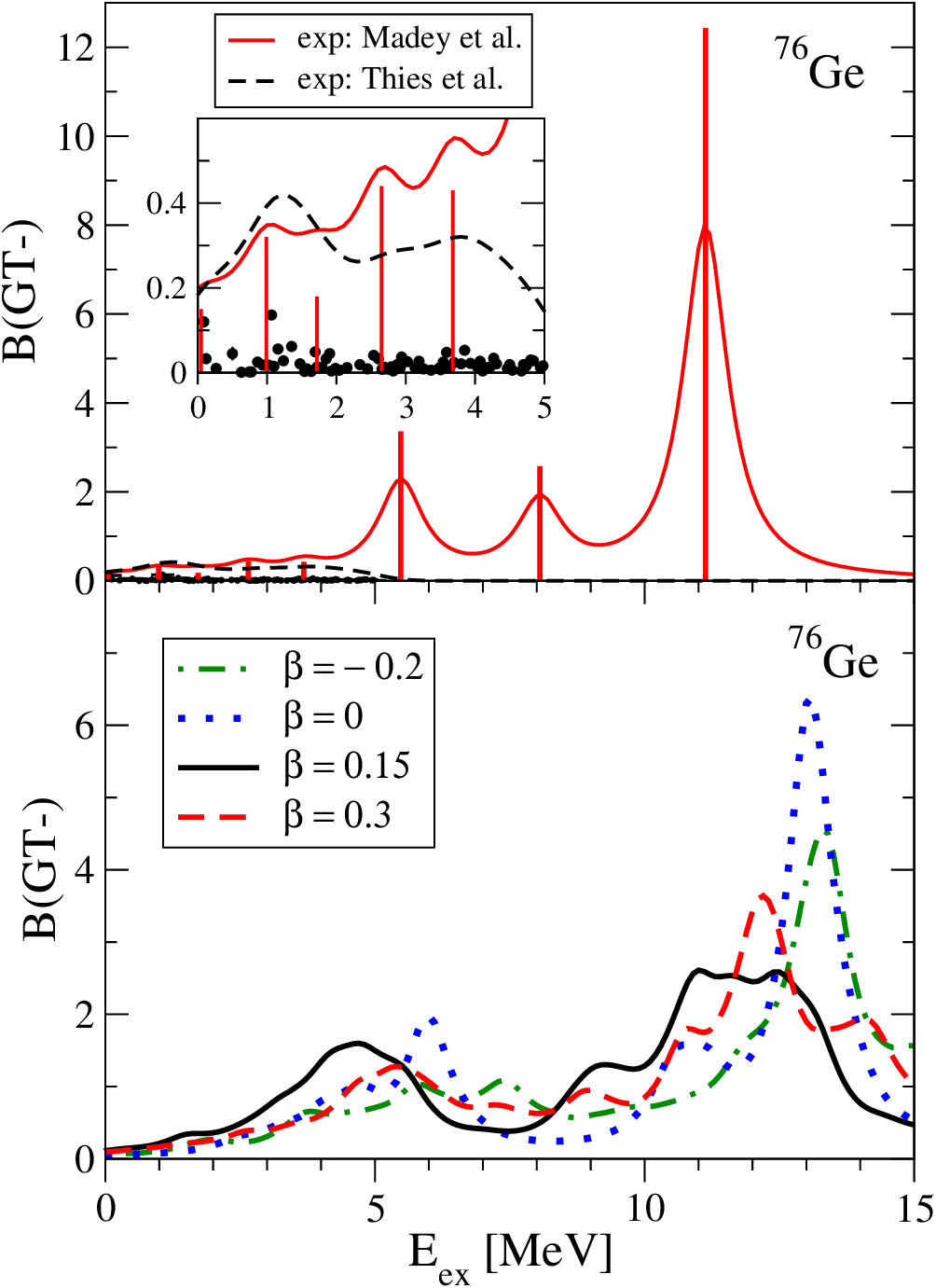}
\caption{(Color online) Gamow-Teller $B({\rm GT}^-)$ strength distributions
in $^{76}$Ge as a function of the excitation energy in the daughter
nucleus. The upper panel shows the data \cite{madey89,thies12} from
different experiments. The lower panel shows calculated QRPA results 
with various quadrupole deformations.}
\label{fig1}
\end{center}
\end{figure}

\begin{figure}[ctb]
\begin{center}
\includegraphics[width=8.5cm]{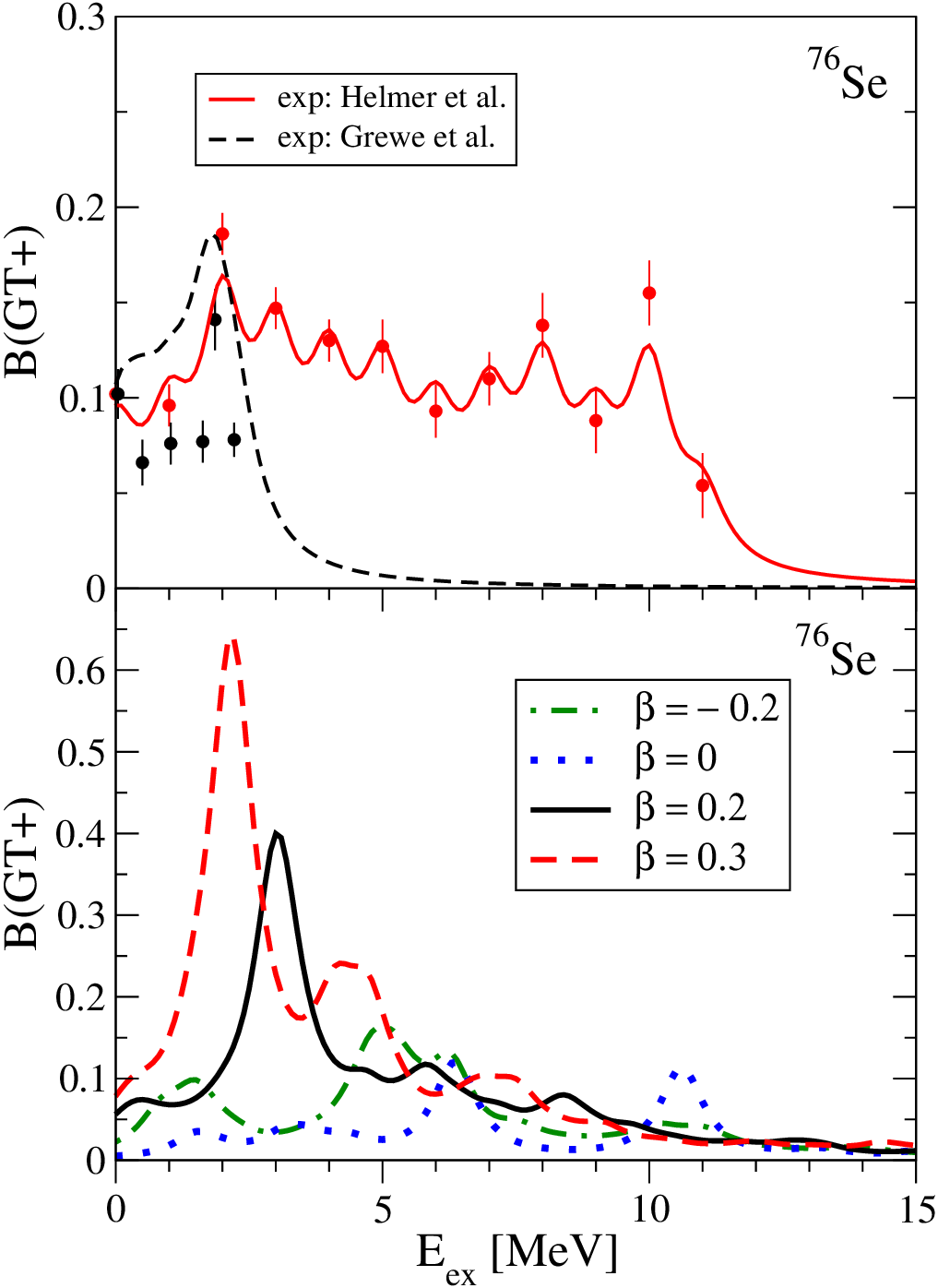}
\caption{(Color online) Same as in Fig.~\ref{fig1}, but for the $B({\rm GT}^+)$
strength in $^{76}$Se. Experimental data are from Refs.~\cite{helmer97,grewe08}.}
\label{fig2}
\end{center}
\end{figure}

\begin{figure*}[ctb]
\begin{center}
\includegraphics[width=16.0cm]{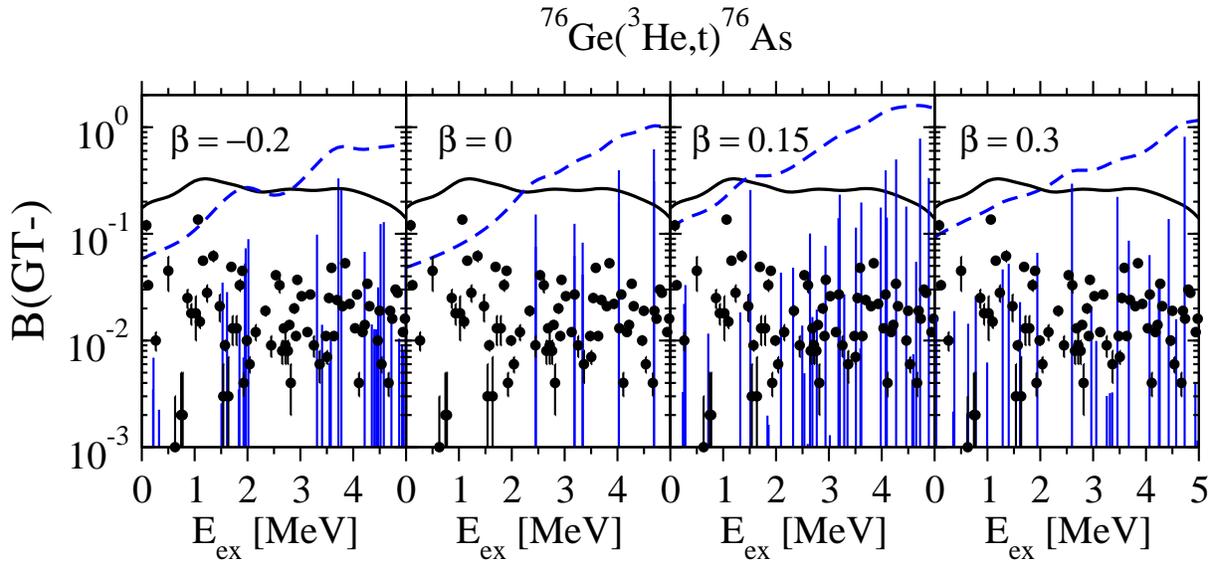}
\caption{(Color online) High resolution $^{76}$Ge($^3$He,t)$^{76}$As
data \cite{thies12} compared to QRPA calculations (dashed and vertical lines) 
with various quadrupole deformations.}
\label{fig3}
\end{center}
\end{figure*}

\begin{figure*}[ctb]
\begin{center}
\includegraphics[width=16.0cm]{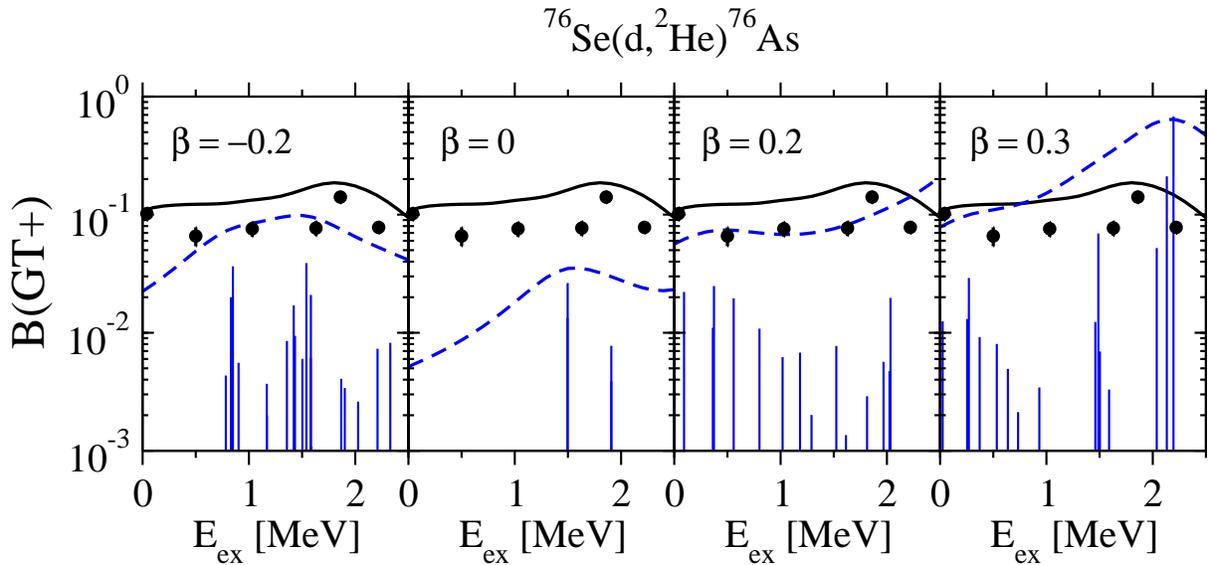}
\caption{(Color online) Same as in Fig.~\ref{fig3}, but for 
$^{76}$Se$(d,^2$He)$^{76}$As data \cite{grewe08}. }
\label{fig4}
\end{center}
\end{figure*}

\begin{figure}[ctb]
\begin{center}
\includegraphics[width=8.5cm]{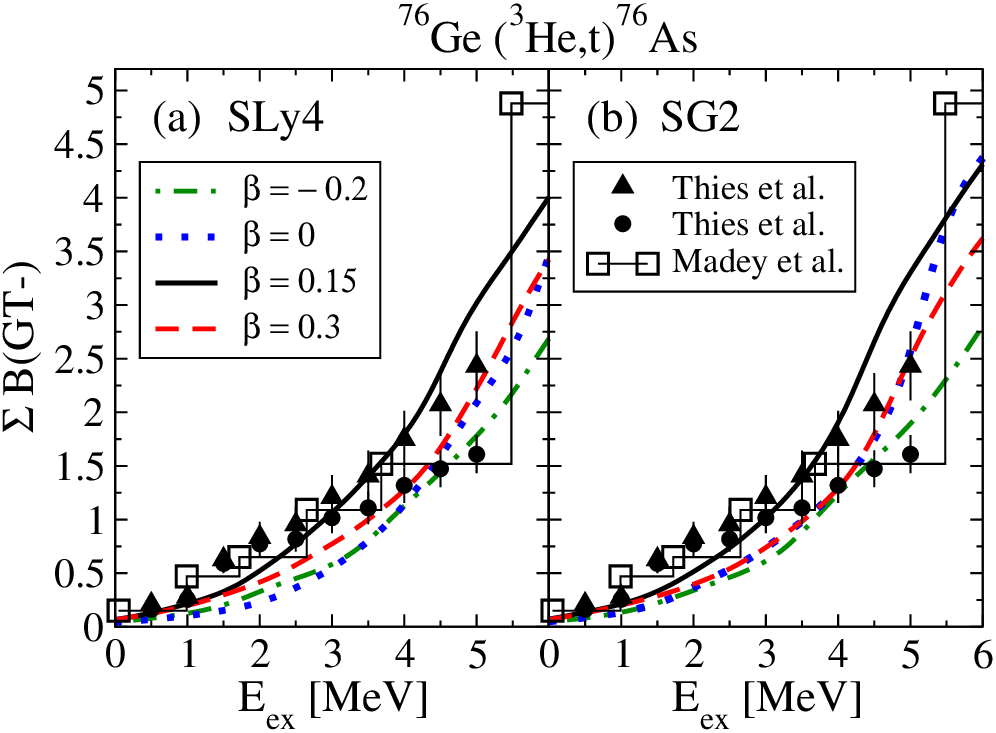}
\caption{(Color online) Accumulated $B({\rm GT}^-)$ of $^{76}$Ge as a 
function of the excitation energy in the daughter nucleus. 
The data \cite{thies12,madey89} are compared with theoretical calculations 
obtained with different quadrupole deformations. }
\label{fig5}
\end{center}
\end{figure}

\begin{figure}[ctb]
\begin{center}
\includegraphics[width=8.5cm]{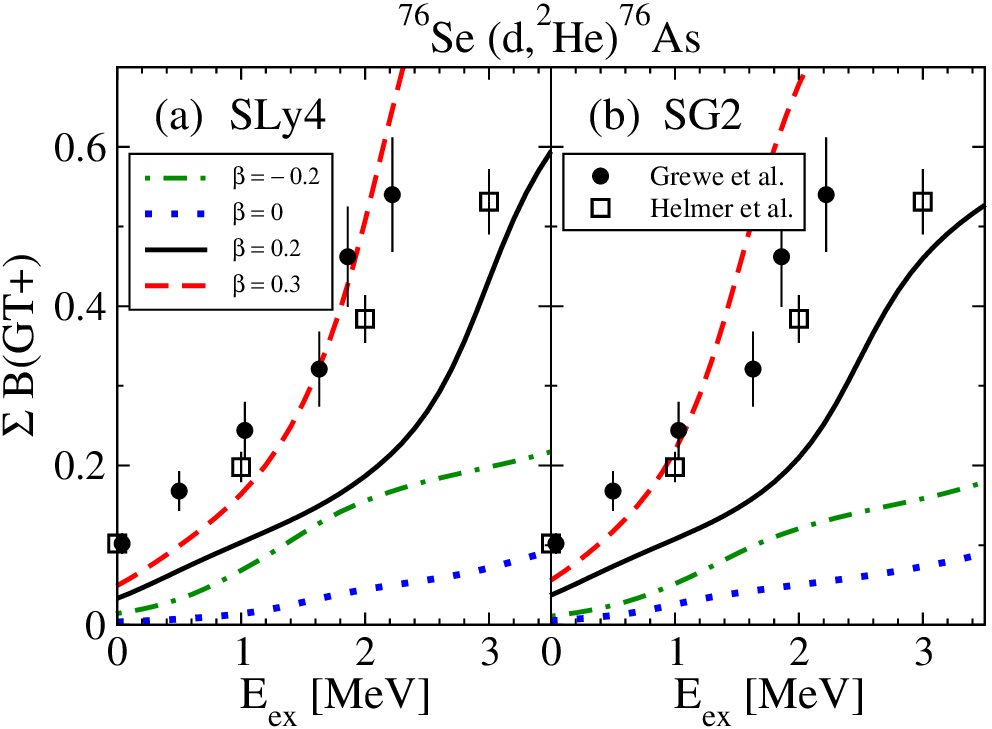}
\caption{(Color online) Same as in Fig.~\ref{fig5}, but for the 
$B({\rm GT}^+)$ of  $^{76}$Se.
Data are from Refs.~\cite{grewe08,helmer97}.  }
\label{fig6}
\end{center}
\end{figure}

\begin{figure}[ctb]
\begin{center}
\includegraphics[width=8.5cm]{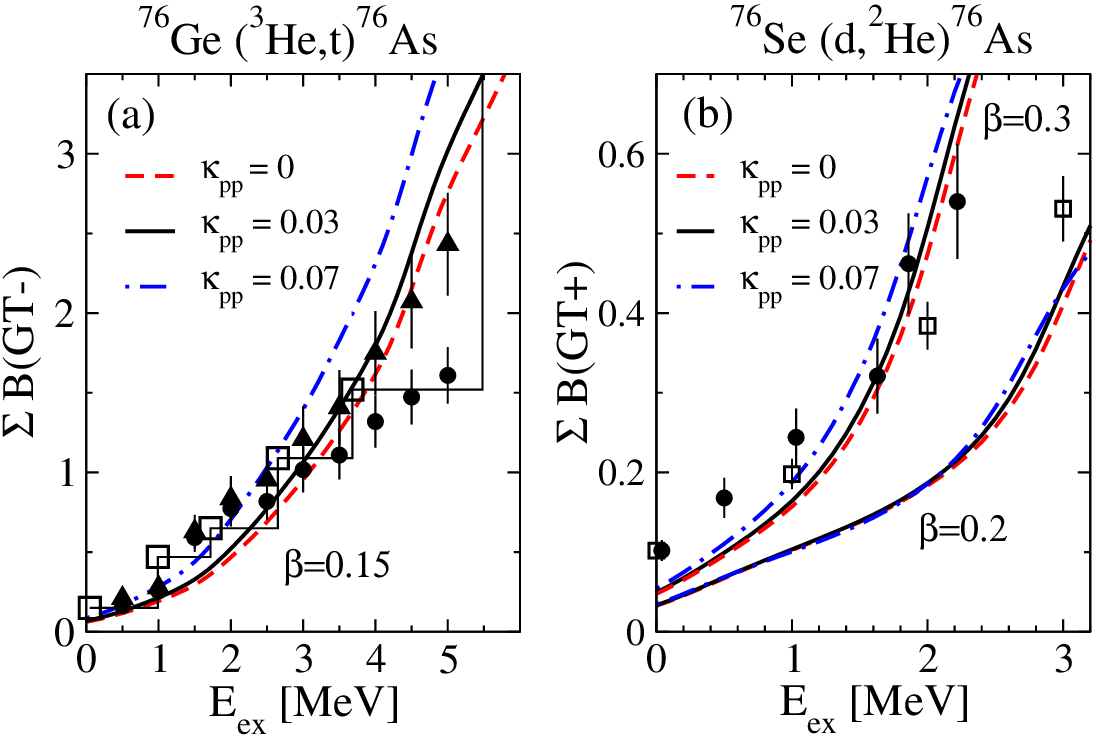}
\caption{(Color online)
Accumulated (a) $B({\rm GT}^-)$ and (b) $B({\rm GT}^+)$ as a 
function of the excitation energy in the daughter nucleus. 
The results correspond to the Skyrme force SLy4 for various values of
the coupling strength $\kappa_{pp}^{GT}$ [MeV].
Experimental points in (a) and (b) are as in Figs.~\ref{fig5} and \ref{fig6},
respectively. }
\label{fig7}
\end{center}
\end{figure}

\begin{figure}[ctb]
\begin{center}
\includegraphics[width=8.5cm]{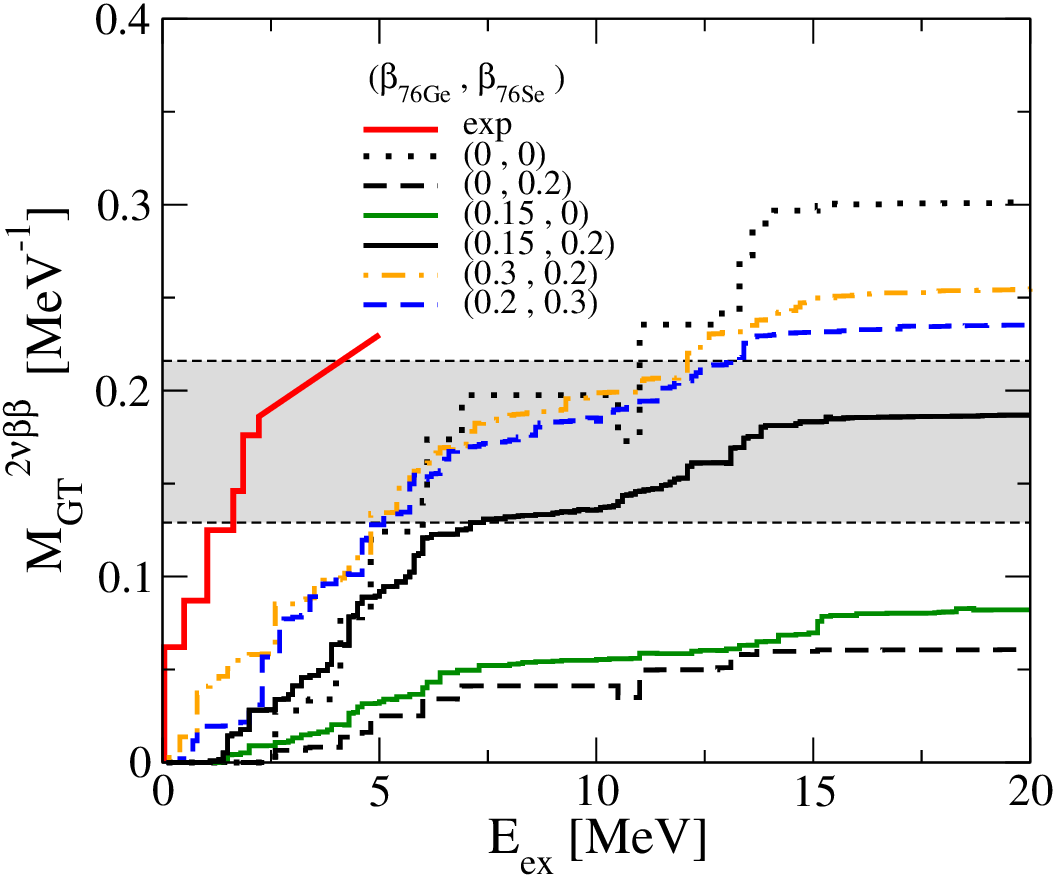}
\caption{(Color online) Experimental running sum \cite{thies12} of the 
nuclear matrix element for the $2\nu\beta\beta$ decay in $^{76}$Ge as a
function of the intermediate excitation energy in $^{76}$As, compared with
results from calculations using different quadrupole deformations for 
parent and daughter nuclei. The shaded area indicates the experimental 
range extracted from the experimental half-life using bare and quenched 
$g_A$ factors.}
\label{fig8}
\end{center}
\end{figure}

\begin{figure}[ctb]
\begin{center}
\includegraphics[width=8.5cm]{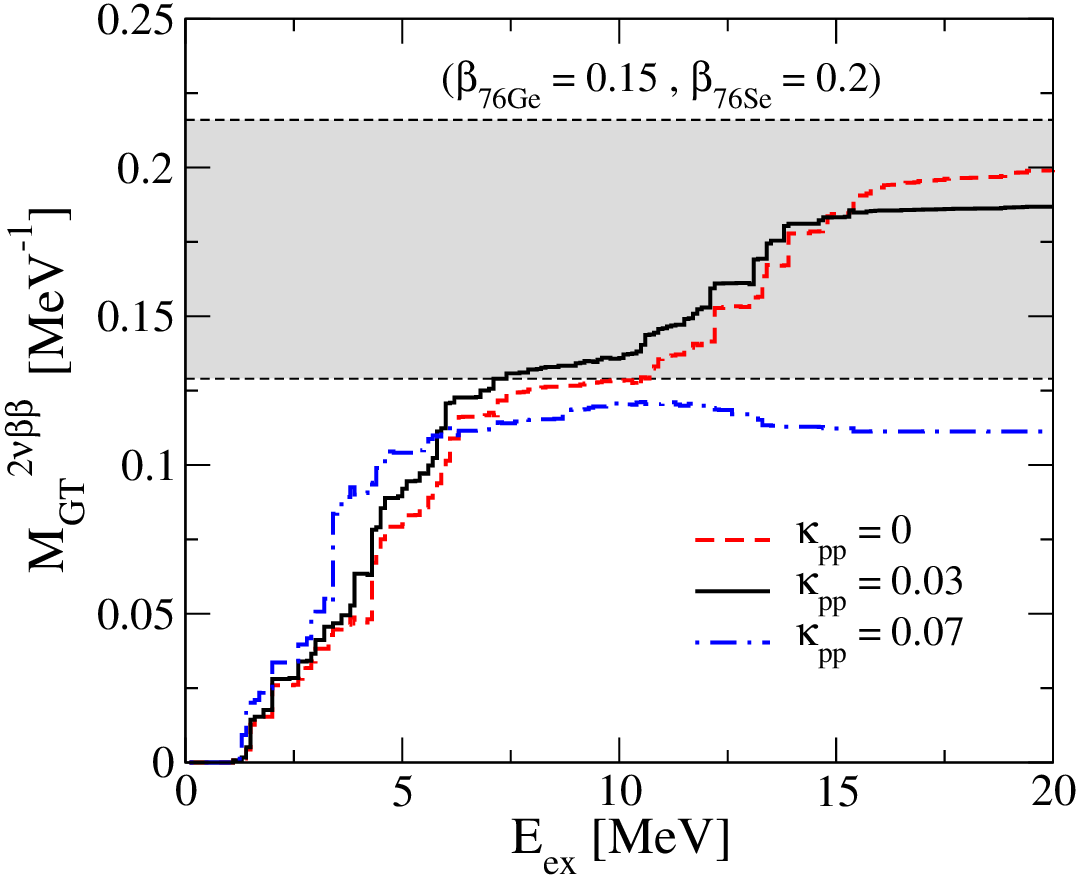}
\caption{(Color online) Same running sums as in Fig.~\ref{fig8}, but
for different values of the coupling strength $\kappa_{pp}^{GT}$ [MeV].}
\label{fig9}
\end{center}
\end{figure}

\begin{figure}[ctb]
\begin{center}
\includegraphics[width=8.5cm]{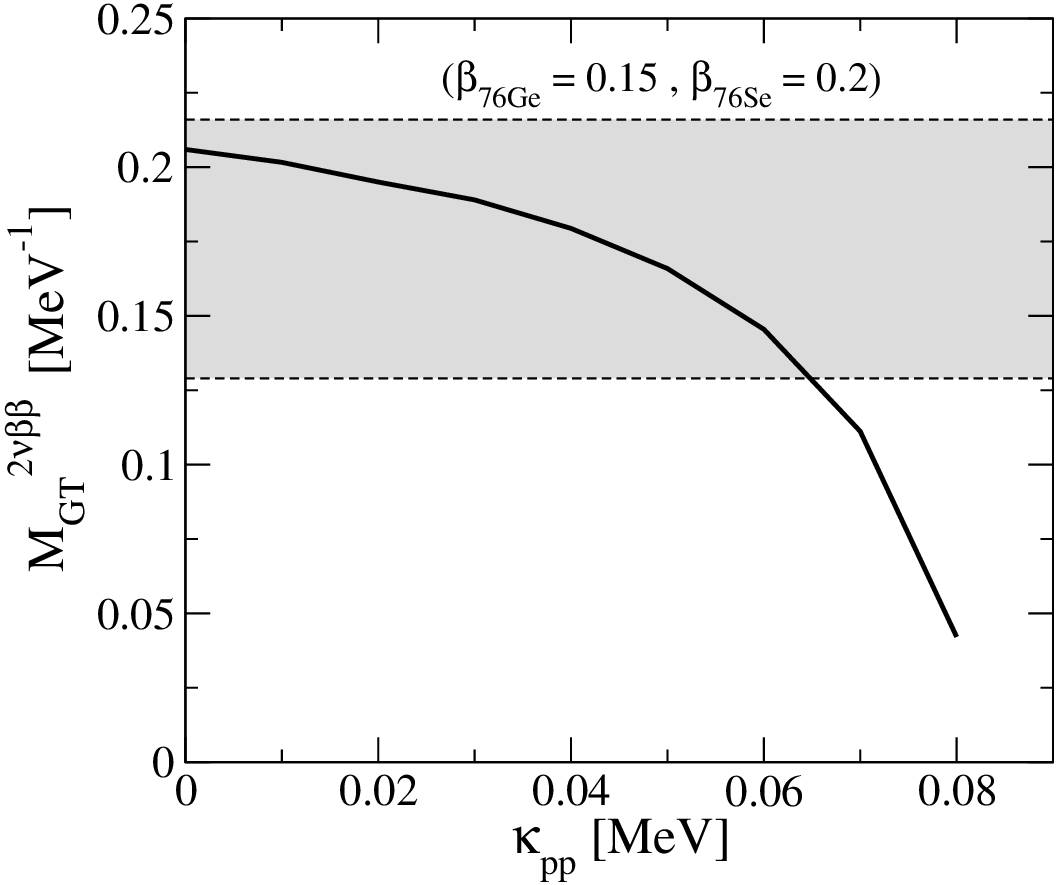}
\caption{Nuclear matrix element for the $2\nu\beta\beta$ decay in $^{76}$Ge 
as a function of the coupling strength $\kappa_{pp}^{GT}$.
The shaded area has the same meaning as in Fig.~\ref{fig8}.}
\label{fig10}
\end{center}
\end{figure}

In this section, we first discuss the global ground-state properties of 
$^{76}$Ge and $^{76}$Se. In Table~\ref{rqb} we compare with experiment our 
results from microscopic SLy4 HF+BCS calculations corresponding to charge 
root mean square radii $r_c$, quadrupole moments $Q_p$, and quadrupole 
deformation parameters $\beta$.
The experimental values for charge radii from Ref.~\cite{radiiexp} are
well reproduced in our calculations.
The experimental intrinsic quadrupole moments quoted correspond to values 
extracted from Coulomb excitation reorientation methods \cite{stone05}
and from electric quadrupole transition probabilities B(E2) \cite{raman01}.
Our calculations produce quadrupole deformations within those experimental
values.

Besides the selfconsistent solution that gives us the energy minimum
and the nuclear shapes at equilibrium, we are also interested in the
behavior of the energy as a function of the deformation. These energy
curves are obtained by performing constrained HF+BCS calculations 
\cite{flocard73}, where the HF energy is minimized under the constraint
of keeping the nuclear deformation fixed.
The energy curves for $^{76}$Ge and $^{76}$Se exhibit two local minima at
oblate and prolate shapes that are practically symmetric with very
low energy barriers. The profiles
of these curves are very shallow and roughly have the same energy for 
quadrupole deformations from $\beta=-0.2$ up to $\beta=0.2$.
These results are in agreement with those from similar calculations
using the Gogny-D1S finite-range effective interaction \cite{webpage}.
Taking into account these characteristics, with oblate, spherical and
prolate shapes having practically the same energy and where small changes
in the calculation can lead to different results for the absolute minimum
associated to the ground state, we have opted to show here results
for the GT strength distributions at various shapes
corresponding to $\beta=-0.2,0,0.15,0.2,0.3$ to study the sensitivity of
the GT strength distributions to the deformation.

\subsection{Gamow-Teller strength distributions}

In the upper plot of Fig.~\ref{fig1} we show the measured GT strength
distribution as a function of the excitation energy of the daughter
nucleus, extracted from the charge-exchange reactions 
$^{76}$Ge$(p,n)^{76}$As \cite{madey89} and 
$^{76}$Ge($^3$He$,t)^{76}$As \cite{thies12}. While the former measurements
were taken at a rather low resolution, the latter experiment was performed
at a high energy resolution of 30 keV. The inset shows a more detailed 
comparison of these two measurements in the overlapping energy range below
5 MeV. The curves correspond to the same distributions obtained from a
folding procedure using 1 MeV width Breit-Wigner functions, so that the 
original discrete spectrum is transformed into a continuous curve. 

In the lower panel we can see the GT strength distributions calculated
within QRPA for various deformation parameters using the same folding
procedure. Taking as a reference the solid black line ($\beta=0.15$), 
we can see that the main characteristics observed, such as the location 
and strength of the GT resonance are qualitatively reproduced. In 
Table~\ref{gtge} we can see a comparison between measured and calculated 
GT strengths with $\beta=0.15$
in different energy regions. In the low energy region up to 
5~MeV, we have two different evaluations of the GT strength measured
in Ref.~\cite{thies12} that compare well with the old measurements of 
Ref.~\cite{madey89}. Our calculations overestimate clearly these measurements
in this range of energy. Nevertheless, if we consider the range of
energy up to 7~MeV, then the calculations produce comparable results. 
This is also true when we compare the strength contained between 7 and 
10~MeV, where another peak is observed, as well as when we compare the
strength contained between 10 and 20~MeV, where the GT resonance is 
located. The total strength in the whole energy range is also well 
described. Calculations with other deformations produce peaks that are
displaced, but qualitatively they are similar with one broad peak
centered at about 5~MeV and another one centered beyond 10~MeV.

Similarly, we can see in Fig.~\ref{fig2} the corresponding results for
the charge-exchange reactions $^{76}$Se$(n,p)^{76}$As \cite{helmer97}
and $^{76}$Ge$(d,^2$He)$^{76}$As \cite{grewe08} with an improved energy 
resolution of 120 keV. The curves represent again folded distributions. 
The lower panel shows the QRPA results for different values of quadrupole
deformations. We see in this case an increased sensitivity to deformation 
in the low energy region where a first peak carrying most of the strength
is particularly enhanced for prolate deformations.
In any case, we should keep in mind that in the case of $B({\rm GT}^+)$ 
the total strength is much lower than in the case of $B({\rm GT}^-)$, as 
it must be according to the Ikeda sum rule for $N>Z$ nuclei.
Taking in this case $\beta=0.2$ (solid black line) as a reference, we can 
see in Table~\ref{gtse} the strength contained in various energy ranges
compared to experiment. The GT strength contained below 2~MeV is largest
for the most recent data of Ref.~\cite{grewe08}. However, as in the case
of $^{76}$Ge, for energies below 5~MeV both sets of data and the theoretical
results are comparable. The total strength contained in the whole energy
range is also well reproduced by the calculations.

This comparison tells us that the global behavior of the GT strength
distribution as a whole is well reproduced in our calculations for
reasonable values of the coupling constants of the residual interaction 
and for deformation parameters compatible with experiment.

In the next two figures, Figs.~\ref{fig3} and \ref{fig4}, we can see
the comparison between the QRPA results for various deformations (blue lines)
and the high resolution measurements for $^{76}$Ge and $^{76}$Se, respectively. 
In both figures we show the folded measured (solid lines) and calculated 
(dashed lines) distributions for a better comparison.

In the case of $^{76}$Ge we can see that we get systematically less (more) 
strength than experimentally observed below (above) an excitation
energy of about 2 MeV. This is true for any deformation and the total
GT strength in this region is somewhat compensated.
On the other hand, it is clear that the highly fragmented strength
measured is only qualitatively reproduced when prolate deformations of 
about $\beta\sim 0.15$ are considered. In particular, the spherical
strength distribution shows a concentration of the strength in a few 
peaks at variance with experiment.
The strength distributions in Fig.~\ref{fig4} corresponding to $^{76}$Se
show also the characteristic fragmentation due to deformation that agrees
better with the experiment. In this case the total calculated strength below
this small range of energy is lower than experiment except for large
deformations that accumulate strength around 2 MeV (see Table~\ref{gtse}).

In Figs.~\ref{fig5} and \ref{fig6} we show the accumulated GT strength 
[$\sum B({\rm GT})(E_{ex})=\sum_{E<E_{ex}}B({\rm GT})(E)$] for $^{76}$Ge and 
$^{76}$Se, respectively. They are calculated from the folded distributions 
and correspond to QRPA calculations with various deformations using two
different Skyrme forces to appreciate the sensitivity of the results
to different parametrizations of the effective Skyrme interaction used. 
In the left panels (a) we have the results from SLy4, whereas in the right
panels (b) we have the results from the Skyrme interaction SG2 \cite{giai}. 
The latter has been successfully tested against spin and isospin excitations 
in spherical \cite{giai} and deformed nuclei \cite{sarriguren,sarrnoj}.
As we can see from these figures, the tendencies are very similar and the
small discrepancies are only quantitative. This type of plot is very
useful because it shows at the same time both the detailed structure
of the strength distribution and the global behavior in terms of total 
strength contained at each excitation energy. In Fig.~\ref{fig5} the 
calculations for the accumulated $B({\rm GT}^-)$ strength in $^{76}$Ge are 
compared to the data from $^{76}$Ge($^3$He$,t)^{76}$As \cite{thies12} 
(circles) and with the data extracted over 0.5 MeV energy bins (triangles) 
that contain the influence of the tail of the GT resonance \cite{thies12}. 
We also show for comparison the data from $(p,n)$ reactions \cite{madey89} 
(open squares).
We can see that the measured strength is systematically underestimated
at low excitation energy, but beyond 4~MeV the tendency is the opposite.
The data are best reproduced by the calculations with a quadrupole 
deformation $\beta=0.15$.
In Fig.~\ref{fig6} we show the calculations for the $B({\rm GT}^+)$
strength in $^{76}$Se for various deformations. They are compared with
data from the $^{76}$Se$(d,^2$He)$^{76}$As \cite{grewe08} (dots) and 
$(n,p)$ reactions \cite{helmer97} (open squares).
In this case the results from spherical and oblate shapes clearly 
underestimate the data. Prolate deformations produce much more strength
and follow better the observed trend.

Finally, it should also be mentioned that the strength of the residual
proton-neutron interaction in the $pp$ channel ($\kappa_{pp}^{GT}$) has 
been shown to play an important role to describe both the GT strength 
distributions \cite{hir,sarri_pp,j_hirsch} and the $2\nu\beta\beta$ 
nuclear matrix elements \cite{muto88,simkovic04,yousef09,alvarez04}. 
To demonstrate the sensitivity of the GT strength distributions to
this parameter, we show in Fig.~\ref{fig7} the accumulated (a)
$B({\rm GT}^-)$ and (b) $B({\rm GT}^+)$ for several values of the 
coupling strength $\kappa_{pp}^{GT}$ [MeV].
The results correspond to the Skyrme force SLy4 for various values of
the deformation parameters close to the selfconsistent ones and to the
experiment. We observe a dispersion of the results characterized by a 
larger accumulation of the strength at lower energies as the value of 
$\kappa_{pp}^{GT}$ increases. A similar tendency is found for other values 
of the deformation parameter. In the case of $B({\rm GT}^-)$, the spread 
in the profile produced by $\kappa_{pp}^{GT}$ is comparable to the spread 
produced by the deformation (see Fig.~\ref{fig5}(a)). On the other hand, 
the effect of $\kappa_{pp}^{GT}$ in the accumulated $B({\rm GT}^+)$ at low 
energies is much smaller than the effect produced by the deformation (see 
Fig.~\ref{fig6}(a)).

\subsection{Two-neutrino double-beta decay}

In this subsection we evaluate the $2\nu\beta\beta$ matrix elements for
the decay of $^{76}$Ge and compare them with the experimental information
extracted from both the measured half-life of the process \cite{barabash10}
and the running sum, as extracted directly from the high resolution 
measurements \cite{thies12,grewe08}.

In Fig.~\ref{fig8} we can see the running sum of the $2\nu\beta\beta$
nuclear matrix element for various combinations of parent and daughter
deformations as a function of the excitation energy of the intermediate
nucleus $^{76}$As. The shaded area indicates the experimental range for
the matrix element $M_{GT}^{2\nu\beta\beta}$ extracted from the experimental 
half-life using bare and quenched $g_A$ factors.
Also included (red solid line) is the experimental running sum up to
5~MeV derived following the procedure explained in Ref.~\cite{thies12}
using the GT strength distributions from \cite{thies12} and \cite{grewe08}.

The spherical description of these nuclei produce a very large matrix
element that overestimates the experiment. The matrix elements are
reduced when different deformations from parent and daughter are
considered. In particular, the combination of deformation parameters
$\beta (^{76}$Ge$)=0.15$ and  $\beta (^{76}$Se$)=0.2$ (solid black line) 
produce optimal results reproducing the experiment. When the deformations
are very different, the matrix element becomes very small as compared to
the experiment.

It should also be noted that all the calculations lie below the
experimental running sum of Ref.~\cite{thies12} that reaches the
experimental value from the half-life already at 2~MeV.
As pointed out by the authors in that reference, the consequence of this
is that any further contributions must be very small or must cancel each
other. However, one should also take into account that the construction
of the running sums in \cite{thies12} was based on summing the products of 
GT$^-$ and GT$^+$ matrix elements accumulated in different overlapping
energy windows. This was done because the different resolutions of the
experiments on the two branches and the lack of correlation between the 
GT$^-$ and GT$^+$ strengths do not allow a one-to-one correspondence 
between the states in $^{76}$As excited from $^{76}$Ge and $^{76}$Se.
Then, this procedure provides an upper limit of the actual running sum 
because the product of accumulated strengths is always larger than the 
sum of one-to-one products. 
In this sense, our calculations for the running sums are consistently 
below the experimental running sum extracted in that way.

The dependence of the running sum with the coupling strength of the 
proton-neutron residual interaction in the $pp$ channel ($\kappa_{pp}^{GT}$) 
is shown in Fig.~\ref{fig9}, using the deformations $\beta (^{76}$Ge)=0.15 
and $\beta (^{76}$Se)=0.20 for the parent and daughter nuclei, respectively.
The concentration of the strength at low energies produced with higher
values of $\kappa_{pp}^{GT}$ makes the $2\nu\beta\beta$ nuclear matrix 
element increase at low energies with increasing values of 
$\kappa_{pp}^{GT}$, but nevertheless, this increase is not enough
to reproduce the experimental running sum extracted in Ref.~\cite{thies12}.
When the running sum exhausts, the final matrix element decreases with 
increasing values of $\kappa_{pp}^{GT}$. This effect can be better appreciated 
in Fig.~\ref{fig10}, where $M_{GT}^{2\nu\beta\beta}$ is plotted as a function of 
$\kappa_{pp}^{GT}$.

\section{Summary and Conclusions}
\label{sec:summary}

In this work we have studied the GT strength distributions in the daughter
nucleus $^{76}$As reached from both $^{76}$Ge and $^{76}$Se double-$\beta$ 
decay partners. Calculations from a deformed QRPA approach with $ph$ and
$pp$ residual interactions based on a selfconsistent Skyrme Hartree-Fock
mean field with pairing correlations are compared with data from $(p,n)$
and $(n,p)$ charge-exchange reactions and their associated high resolution 
($^3$He$,t)$ and $(d,^2$He) reactions.

Using quadrupole deformations compatible with the equilibrium shapes
obtained from the SLy4 interaction, which lie within the experimental
values, we obtain reasonable agreement with experiment in both single-beta 
branches, GT$^-$ in $^{76}$Ge and GT$^+$ in $^{76}$Se, as measured in 
Refs.~\cite{madey89,thies12} and \cite{helmer97,grewe08}, respectively,
as well as with the nuclear matrix element of the $2\nu\beta\beta$ 
process, extracted from the experimental half-life \cite{barabash10}.
The deformed QRPA approach used in this work is able to reproduce the
gross features of the GT strength distributions, but fails to account 
for a detailed description of the high resolution data in the low 
excitation energy.
Although a fine spectroscopy is beyond the scope of this HF+BCS+QRPA 
approach, where the use of universal effective Skyrme interactions 
(SLy4 in this work) intended to describe spherical and deformed nuclei 
all along the nuclear chart prevents reproducing accurately local details,
the global performance of this approach is very reasonable.
It is also worth mentioning that the agreement with experiment is
optimum for the most reasonable choices of deformations and strengths
of the residual interaction.

It has been shown that nuclear deformation plays a significant role in
understanding the GT strength distribution, as well as in understanding the 
$2\nu\beta\beta$ process, where differences between parent and daughter 
nuclear deformations introduce a reduction factor in the nuclear matrix
elements that finally determines the $2\nu\beta\beta$ half lives.

The role of the coupling strength of the proton-neutron residual
interaction in the $pp$ channel ($\kappa_{pp}^{GT}$) to describe GT
strength distributions and $2\nu\beta\beta$ nuclear matrix elements has
been studied. It has been shown that, within a range of reasonable
values of $\kappa_{pp}^{GT}$, its effect on the GT strength distributions
at low energy is comparable or smaller than the effect produced by the
deformation. It has also an effect on the $2\nu\beta\beta$ matrix elements,
especially on the total matrix element that finally determines the
$2\nu\beta\beta$ half-life.

\section*{Acknowledgments \label{sec:acknowledge}}
\addcontentsline{toc}{chapter}{Acknowledgments}

I am grateful to D. Frekers for access to the paper in 
Ref. \cite{thies12} prior to its publication.
This work was supported in part by MINECO (Spain) under Research Grant
No.~FIS2011--23565 and by Consolider-Ingenio 2010 Programs CPAN 
CSD2007-00042.

\end{document}